# Monopoles and hadron spectrum in quenched QCD

Tsuneo Suzuki [a], Shun-ichi Kitahara [a], Tsuyoshi Okude [a], Fumiyoshi Shoji [a], Kazuya Moroda [a] and Osamu Miyamura [b]

[a]Department of Physics, Kanazawa University, Kanazawa 920-11, Japan

[b]Department of Physics, Hiroshima University, Higashi Hiroshima 739, Japan

We report the preliminary results of the studies of hadron spectrum under the background of abelian and monopole gauge fields in quenched Wilson SU(3) QCD. Abelian gauge fields alone reproduce the same chiral limit as in the full case. Critical hopping parameter $\kappa_c$ and $m_\rho$ are the same in both cases. We need more time to get a definite result in the case of monopole background. The photon contribution do not produce any mass gap in the chiral limit ($\kappa = \kappa_c \sim 0.17$). The behavior is similar to those in the free photon case for $\kappa_c = 0.125$.

## 1. Introduction

Recent Monte-Carlo simulations suggest that abelian monopoles after abelian projection are a key quantity of confinement mechanism in QCD [1–3]. Especially the abelian projection in the maxumally abelian (MA) gauge is interesting. Then mass generation from scale invariant QCD may be explained by the abelian monopoles alone. The expectation is supported by the preliminary data of recent simulations in quenched $SU(2)$ QCD with Kogut-Susskind fermions[4]. The purpose of this study is to report preliminary results of the studies of hadron spectrum under the background of abelian and monopole gauge fields in quenched SU(3) QCD with Wilson fermions.

## 2. Maximally abelian gauge in SU(3) QCD

We adopt the usual $SU(3)$ Wilson action. The MA gauge is given by performing a gauge transformation, $\widetilde{U}(s,\mu) = V(s)U(s,\mu)V^{-1}(s+\hat{\mu})$, such that

$$R = \sum_{s,\mu} (|\widetilde{U}_{11}(s,\mu)|^2 + |\widetilde{U}_{22}(s,\mu)|^2 + |\widetilde{U}_{33}(s,\mu)|^2)$$

is maximized. Then a matrix

$$\sum_{\mu,a} [\widetilde{U}(s,\mu)\Lambda_a \widetilde{U}^\dagger(s,\mu) + \widetilde{U}^\dagger(s-\hat{\mu},\mu)\Lambda_a \widetilde{U}(s-\hat{\mu},\mu), \Lambda_a]$$

is diagonalized. Here

$$\Lambda_1 = \mathrm{diag}(1,-1,0), \Lambda_2 = \mathrm{diag}(-1,0,1),$$

$$\Lambda_3 = \mathrm{diag}(0,1,-1).$$

After the gauge fixing is over, we can extract an abelian link field,

$$\widetilde{U}(s,\mu) = A(s,\mu)u(s,\mu),$$

where

$$u(s,\mu) = \mathrm{diag}(e^{i\theta^{(1)}(s,\mu)}, e^{i\theta^{(2)}(s,\mu)}, e^{i\theta^{(3)}(s,\mu)})$$

$$\theta^{(i)}(s,\mu) = \arg(\widetilde{U}_{ii}(s,\mu)) - \frac{1}{3}\phi(s,\mu)$$

$$\phi(s,\mu) = \sum_i \arg(\widetilde{U}_{ii}(s,\mu))|_{\mathrm{mod}\,2\pi}.$$

$u(s,\mu)$ is a diagonal abelian gauge field.

## 3. Abelian gauge fields from monopole Dirac strings

A plaquette variable is given by $f_{\mu\nu}^{(i)}(s) = \partial_\mu \theta_\nu^{(i)}(s) - \partial_\nu \theta_\mu^{(i)}(s)$, where

$$\theta_\mu^{(i)}(s) = -\sum_{s'} D(s-s')[\partial_\nu' f_{\nu\mu}^{(i)}(s') + \partial_\mu(\partial_\nu' \theta_\nu^{(i)}(s'))]$$

and $D(s-s')$ is the lattice Coulomb propagator.

Define abelian gauge filed as

$$\theta_\mu^{(i)}(s) = -\sum_{s'} D(s-s')\partial_\nu' f_{\nu\mu}^{(i)}(s').$$

The gauge fields automatically satisfy the Landau gauge condition.

Extract the Dirac strings from the field strength, satisfying $\sum_i n_{\mu\nu}^{(i)}(s) = 0$:

$$f_{\mu\nu}^{(i)}(s) = \bar{f}_{\mu\nu}^{(i)}(s) + 2\pi n_{\mu\nu}^{(i)}(s),$$

Then an abelian gauge field from Dirac string is

$$\theta_\mu^{Ds(i)}(s) = -2\pi \sum_{s'} D(s-s')\partial'_\nu n^{(i)}_{\nu\mu}(s'),$$

whereas an abelian gauge field from the photon is

$$\theta_\mu^{Ph(i)}(s) = -\sum_{s'} D(s-s')\partial'_\nu \bar{f}^{(i)}_{\nu\mu}(s').$$

Abelian link fields from Dirac string and photon are constructed as

$$u^{Ds}(s,\mu) = \mathrm{diag}(e^{i\theta_\mu^{Ds(1)}(s)}, e^{i\theta_\mu^{Ds(2)}(s)}, e^{i\theta_\mu^{Ds(3)}(s)}),$$
$$u^{Ph}(s,\mu) = \mathrm{diag}(e^{i\theta_\mu^{Ph(1)}(s)}, e^{i\theta_\mu^{Ph(2)}(s)}, e^{i\theta_\mu^{Ph(3)}(s)}).$$

## 4. Inverse of Wilson quark matrix

Now let us evaluate the inverse of Wilson fermion matrix

$$1 - \kappa \sum_\mu \; [ \; (1-\gamma_\mu)U(s,\mu)\delta_{s+\hat{\mu},s'}$$
$$+ \; (1+\gamma_\mu)U^\dagger(s-\hat{\mu},\mu)\delta_{s-\hat{\mu},s'}]$$

for

- $U(s,\mu)$ (full)
- $u(s,\mu)$ (abelian)
- $u^{Ds}(s,\mu)$ (Dirac string)
- $u^{Ph}(s,\mu)$ (photon),

and then measure hadron mass spectrum.

The simulations are performed as follows:

- Our calculation is usually done on a 8PE's system of a vector parallel supercomputer Fujitsu VPP500. Lattice size adopted is $16^3 \times 32$. $\beta = 5.7$.

- Antiperiodic (Periodic) boundary conditions in the time (space) direction.

- Number of sweeps for thermalization is 3000.

- Number of sweeps to get independent configurations is 1000.

- Number of configurations for average is 20.

- The criterion of the MA gauge condition is $\sum |\text{off diagonal part of } X| < 10^{-7}$ where ($X$ is the operator to be diagonalized.) Over-relaxation method with $\omega = 1.93$ is used.

- The criterion of the solution of Wilson matrix inversion is $|\text{residue}| < 10^{-5}$. The hopping parameter expansion and the CG method with red-black preconditioning is used.

- The mass fitting function is $c\cosh(m(t-N_t/2))$ and the mass fitting range is $4 \leq N_t \leq 28$.

The CPU time, the number of terms of the hopping parameter expansion and the number of CG iterations in a quenched propagator computation are listed in Table 1.

| $SU(3)$: | | | |
|---|---|---|---|
| $\kappa$ | time(sec.) | HPE | CG |
| 0.160 | 230 | 1200 | 240 |
| 0.165 | 551(16PE) | 1200 | 3000 |
| abelian: | | | |
| $\kappa$ | time(sec.) | HPE | CG |
| 0.150 | 482 | 360 | 1560 |
| 0.155 | 910 | 300 | 3600 |
| 0.1575 | 1188 | 120 | 4440 |
| 0.160 | 2074 | 300 | 8040 |
| 0.1625 | 3000(16PE) | 300 | 20400 |
| monopole: | | | |
| $\kappa$ | time(sec.) | HPE | CG |
| 0.135 | 345 | 1200 | 600 |
| 0.140 | 429 | 180 | 1560 |
| 0.1425 | 346(16PE) | 120 | 2160 |

Table 1
Various parameters of the simulations.

## 5. Results

Examples of hadron propagators with abelian and monopole backgrounds are shown in Fig. 1 and Fig. 2.

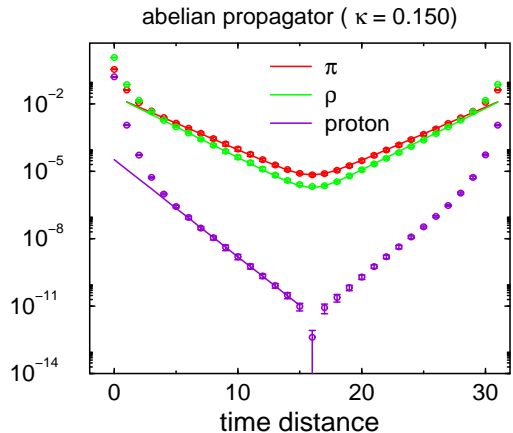

Figure 1. Hadron propagators with abelian background for $\kappa = 0.150$.

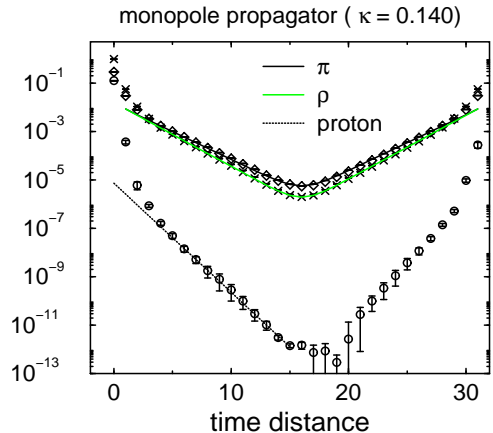

Figure 2. Hadron propagators with monopole background for $\kappa = 0.140$.

The estimated masses of $\pi$, $\rho$ and nucleon are listed in Table 2[5].

The squared masses of $\pi$ and $\rho$ and the mass of nucleon and $\rho$ are plotted versus $1/\kappa$ in Fig. 3 and Fig. 4. Abelian gauge fields alone reproduce the same chiral limit as in the full case. Critical hopping parameter $\kappa_c$ and $m_\rho$ are the same in both cases as expected. The abelian case gives a smaller ratio $m_\pi/m_\rho$ at the same $\kappa$. At present we have obtained the ratio as small as $m_\pi/m_\rho \sim 0.48$ at $\kappa = 0.1625$.

The abelian case gives a smaller ratio $m_p/m_\rho$ at the same $\kappa$. Namely the $O(a)$ correction of this case is smaller in the abelian case. At present we have obtained the ratio as small as $m_p/m_\rho \sim 1.32$ for $m_\rho a = 0.55$ ( at $\kappa = 0.1575$).

We need more time to get the monopole contribution clearly and it is in progress.

Fig. 5 shows the squared masses of $\pi$ and $\rho$ versus $1/\kappa$ under the photon and free backgrounds. The photon contribution do not produce any mass gap in the chiral limit ($\kappa = \kappa_c \sim 0.17$). The behavior is similar to those in the free photon case, although $\kappa_c = 0.125$ in the free case.

In summary, the above preliminary data suggest abelian and also monopole gauge fields are essential for the mass generation of hadrons in QCD. More data are needed to get a definite conclusion and it is under progress. However the data are consistent with the results of recent simulations in quenched $SU(2)$ QCD with Kogut-Susskind fermions[4].

This work is financially supported by JSPS Grant-in Aid for Scientific Research (B) (No.06452028) (T.S.) and (C) (No.07640411) (O.M.).

Table 2
Hadron masses.

| $SU(3)$: | | | |
|---|---|---|---|
| $\kappa$ | $m_\pi$ | $m_\rho$ | $m_N$ |
| 0.160 | 0.689(3) | 0.813(4) | 1.256(18) |
| 0.165 | 0.459(4) | 0.676(1) | 1.045(25) |
| 0.169($\kappa_c$) | 0.0 | 0.543 | — |
| abelian: | | | |
| $\kappa$ | $m_\pi$ | $m_\rho$ | $m_N$ |
| 0.150 | 0.545(6) | 0.630(5) | 1.00(2) |
| 0.155 | 0.442(5) | 0.543(4) | 0.81(3) |
| 0.1575 | 0.364(4) | 0.514(6) | 0.68(7) |
| 0.160 | 0.300(4) | 0.492(9) | — |
| 0.1625 | 0.244(4) | 0.503(23) | — |
| 0.170 ($\kappa_c$) | 0.0 | 0.553 | — |
| monopole: | | | |
| $\kappa$ | $m_\pi$ | $m_\rho$ | $m_N$ |
| 0.135 | 0.671(5) | 0.713(5) | 1.29(7) |
| 0.140 | 0.533(3) | 0.606(3) | 1.04(3) |
| 0.1425 | 0.500(5) | 0.581(6) | 0.91(5) |
| 0.150 | 0.314(6) | 0.442(14) | - |
| 0.1575 | 0.139(18) | 0.247(102) | - |

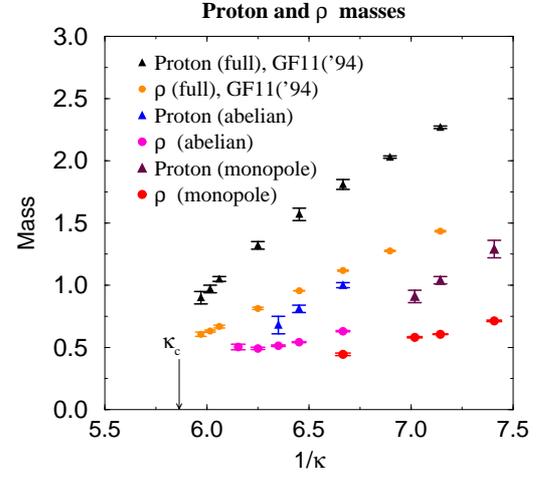

Figure 4. The masses of nucleon and $\rho$ versus $1/\kappa$.

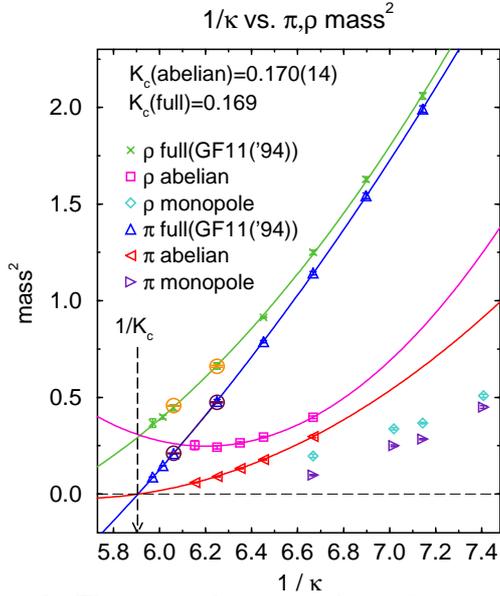

Figure 3. The squared masses of $\pi$ and $\rho$ versus $1/\kappa$.

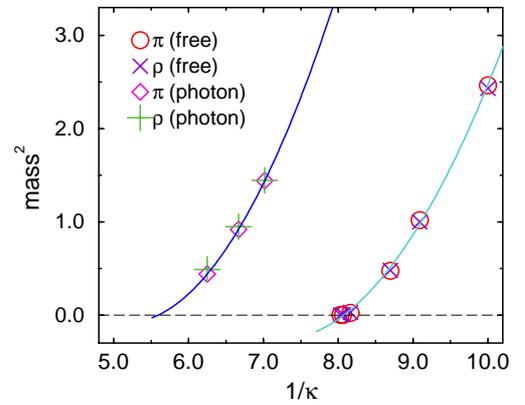

Figure 5. The masses of nucleon and $\rho$ versus $1/\kappa$.

4